\documentclass[aip,cha, reprint,showpacs,showkeys]{revtex4-1}
\usepackage{amssymb, dcolumn,  graphicx, latexsym, amsfonts, bm}

\begin{document}
 
\title{The ion-acoustic turbulence in the skin layer of the inductively coupled plasma}
\author{V. V. Mikhailenko}\email[E-mail: ]{vladimir@pusan.ac.kr}
\affiliation{Plasma Research Center, Pusan National University,  Busan 46241, South Korea.}
\affiliation{BK21 FOUR Information Technology, Pusan National University,  Busan 46241, South Korea.}
\author{V. S. Mikhailenko}\email[E-mail: ]{vsmikhailenko@pusan.ac.kr}
\affiliation{Plasma Research Center, Pusan National University,  Busan 46241, South Korea.}
\author{H. J. Lee}\email[E-mail: ]{haejune@pusan.ac.kr}
\affiliation{Department of Electrical Engineering, Pusan National University, Busan 46241, South Korea.}

\begin{abstract}
The theory of the nonmodal ion-acoustic instability in the skin layer of the inductively coupled plasma (ICP) is developed. This instability has a time -
dependent growth rate and is driven by the current formed in the skin layer by the accelerated motion of electrons relative to ions under the action of the 
ponderomotive force. We found that the development of the ion-acoustic turbulence (IAT) in the skin layer and the scattering of electrons by IAT are primary 
channels of the nonlinear absorption of the RF energy in the skin layer.
\end{abstract}

\maketitle

\section{Introduction}\label{sec1}

The inductively coupled plasma (ICP) is a widespread and dominant plasma source for materials processing applications and operated in the low gas pressure 
regime of a few mTorr range~\cite{Lieberman}. In this regime, the electron mean-free-path is comparable to the characteristic size of plasma devices, 
and the electron collision frequency with neutrals is much less than the driving frequency, $\omega_{0}$, of the applied electromagnetic (EM) wave. 
The absorption of the RF energy by electrons in this regime occurs due to the 
breaking of the phase coherence between the velocity of the individual thermal electron and the spatially nonuniform EM wave with frequency $\omega_{0}$ much 
less than the electron plasma frequency $\omega_{pe}$. The RF electric field in ICPs with such a frequency is non-propagating and localized mainly in the 
skin layer near the plasma boundary adjacent to the RF antenna, where the main interaction of EM fields with plasma electrons takes place. This phenomenon 
involving the spatially inhomogeneous RF fields and nonlocal electron kinetics is well-known in plasma physics as the Landau damping effect. This regime of 
the ICP operation is also named as the regime of the anomalous skin effect\cite{Weibel,Kolobov,Alexandrov} or nonlocal regime. 

The theory of the anomalous skin effect\cite{Weibel,Kolobov,Alexandrov} is based on the solution of the boundary value problem for the combination of the 
linearized Vlasov equation for the electron  distribution function and the Maxwell equations. This linear theory is developed for the limit of the weak RF 
field applied to the plasma boundary for
which the quiver velocity of the plasma electrons in this spatially inhomogeneous RF field was assumed to be negligibly small comparing with the electron 
thermal velocity, $v_{Te}$, and has been neglected. 

It is well known, however, that the relative motion of electrons and ions in a plasma under the action of EM 
waves is the source of the development of numerous parametric and current driven instabilities\cite{Silin,Porkolab,Akhiezer}. The development of these 
instabilities is an alternative channel of the absorption of the EM wave and heating of plasma components, which is not included in the linear theory of the 
anomalous skin effect. 

It was found that the uniformly accelerated motion of electrons occurs in the skin layer, which stems from the ponderomotive force formed by the coupled 
action of the electric and the magnetic component of EM waves \cite{Mikhailenko3}. The kinetic stability theory of the plasma with accelerated electrons
\cite{Mikhailenko3} found that the electrons accelerated in skin layer triggers the current-driven instabilities.  The analytical treatment of the 
instabilities driven by a current with spatially inhomogeneous or time dependent  
current velocity can't be investigated by employing the normal mode analysis, which assumes that the plasma perturbations have a structure 
$\sim \exp\left(i\mathbf{kr}-i\omega t\right)$ of a plane wave.

We employed the nonmodal approach\cite{Mikhailenko3}, which starts with the transformation of the position $\mathbf{r}$ and the velocity $\mathbf{v}$ 
variables of the Vlasov equation for the distribution function $F_{\alpha}\left(\mathbf{v}, \mathbf{r}, t\right)$ of species $\alpha$ (ions or electrons) 
determined in the laboratory frame to the variables $\mathbf{r}_{\alpha}$ and $\mathbf{v}_{\alpha}$ determined in the frame moving with spatially 
inhomogeneous time-dependent velocity $\mathbf{V}_{\alpha}\left(\mathbf{r}, t\right)$. In our approach, this spatially inhomogeneous velocity is determined 
by the Euler equation for the ideal fluid of particles species $\alpha$ 
immersed in the spatially inhomogeneous non-stationary EM field. The Vlasov equation in the moving coordinates  $\mathbf{r}_{\alpha}$ and $\mathbf{v}
_{\alpha}$ has the same form as that in a plasma without an external EM field 
at a finite time interval during which a particle does not move into appreciably different regions of the EM field. The solution of this Euler equation was 
derived in our previous study\cite{Mikhailenko3} for the case of the high frequency RF wave, for which the force by the RF electric field acting 
on electrons in the skin layer prevails over the Lorentz force by the RF magnetic field. It was found that the electron 
velocity $\mathbf{V}_{e}$ in this case is equal to the electron accelerated velocity under the action of the ponderomotive force. 

For the frequency range of the applied RF wave corresponding to the classical skin effect\cite{Kolobov,Alexandrov}, it was revealed\cite{Mikhailenko3} that 
the accelerated electrons leave the skin layer in a short time for the strong RF field for which the ponderomotive current velocity $U$ is larger than the 
electron thermal velocity $v_{Te}$. 
This time appears to be insufficient for the development of the Buneman instability in the skin layer, which excites when the current velocity is larger 
than the electron thermal velocity. It was also found that electrons 
accelerated in the skin layer can trigger the ion-acoustic (IA) instability in the bulk plasma past the skin layer when the quasi-steady electron current 
velocity exceeds the IA velocity $v_{s}=\left(T_{e}/m_{i}\right)^{1/2}$ in this region\cite{Mikhailenko3}. 

In this paper, we apply our nonmodal approach to the theory of the IA instability driven by the weak RF field, for which the accelerated electron current 
velocity in the skin layer is less than the electron thermal velocity, but exceeds the IA velocity. 
Under this condition, the accelerated electrons can develop the IA instability in the skin layer. 
In Sec. \ref{sec2}, we present basic transformations of the system of the Vlasov-Poisson equations, 
employed in the developed theory. We introduce a more general and simple solution of the Euler equation, presented in Appendix, for the electron velocity 
$\mathbf{V}_{e}(\mathbf{r}, t)$ in the decaying EM field without an initial assumption of the stronger electric force than the Lorentz force acting on 
electrons in the skin layer. In Sec. \ref{sec3}, we present the theory of the IA instability in the skin layer driven by the accelerated electrons. The 
nonlinear theory of the IA turbulence in the skin layer is presented in Sec. \ref{sec4} followed by conclusions in Sec. \ref{sec5}.

\section{The nonmodal approach to the theory of the instabilities driven by the accelerated current in the skin layer}\label{sec2}

In this paper, we consider the effect of the relative motion of plasma species on the 
development of the short-scale electrostatic perturbations in the skin layer under the condition of the classical skin effect. The skin effect is classified 
as classical or normal when the frequency $\omega_{0}$ of the RF field belongs to the frequency range\cite{Kolobov,Alexandrov}
\begin{eqnarray}
&\displaystyle
\omega_{pe}\frac{v_{Te}}{c}\ll \omega_{0}\ll \omega_{pe},
\label{1}
\end{eqnarray}
where $\omega_{pe}$ is the electron plasma frequency, and $v_{Te}$ is the electron thermal velocity.   

We consider a plasma occupying region $z\geqslant 0$. The RF antenna which launches the RF wave with a frequency $\omega_{0}$ is assumed to exist to the left 
of the plasma boundary $z=0$. In the frequency range of Eq. (\ref{1}), the electric, $\mathbf{E}_{0}$, and the magnetic, $\mathbf{B}_{0}$ fields of the RF 
wave are exponentially decaying with $z$ and sinusoidally varying with time, 
\begin{eqnarray}
&\displaystyle 
\mathbf{E}_{0}\left(z,t\right)= \mathbf{e}_{y} E_{0y}e^{-\kappa z}\sin \omega_{0}t,
\label{2}
\end{eqnarray}
and 
\begin{eqnarray}
&\displaystyle 
\mathbf{B}_{0}\left(z,t\right)= \mathbf{e}_{x} E_{0y}\frac{c\kappa}{\omega_{0}}e^{-\kappa z}\cos \omega_{0}t ,
\label{3}
\end{eqnarray}
where $\mathbf{E}_{0}$ and $\mathbf{B}_{0}$ satisfy the Faraday’s law, $\partial E_{0}/
\partial z=\partial B_{0}/c\partial t$, and   $\kappa^{-1}=L_{s}$
is the skin depth for the classical skin effect\cite{Kolobov, Alexandrov}, 
\begin{eqnarray}
&\displaystyle L_{s}=\frac{c}{\omega_{pe}}.
\label{4}
\end{eqnarray}
Our theory bases on the Vlasov equations for the velocity distribution functions $F_{\alpha}$ of species $\alpha$ 
($\alpha=e$ for electrons and $\alpha=i$ for ions), 
\begin{eqnarray}
&\displaystyle 
\frac{\partial F_{\alpha}}{\partial t}+\mathbf{v}\cdot\frac{\partial F_{\alpha}}
{\partial\mathbf{r}}+\frac{e_{\alpha}}{m_{\alpha}}\left(\mathbf{E}_{0}
\left(z,t\right)+\frac{1}{c}\left[\mathbf{v}\times\mathbf{B}_{0}\left(z,t \right)\right] \right.
\nonumber
\\ 
&\displaystyle
-\nabla \varphi\left(\mathbf{r},t\right) 
\left)\cdot \frac{\partial F_{\alpha}}{\partial\mathbf{v}}\right. =0,
\label{5}
\end{eqnarray}
and the Poisson equation for the electrostatic potential $\varphi\left(\mathbf{r},t\right)$ of the electrostatic plasma perturbations,
\begin{eqnarray}
&\displaystyle \nabla^2 \varphi\left(\mathbf{r},t\right)=
-4\pi\sum_{\alpha=i,e} e_{\alpha}\int f_{\alpha}\left(\mathbf{v},
\mathbf{r}, t \right)d\textbf {v}_{\alpha}. \label{6}
\end{eqnarray}
Here,  $f_{\alpha}$ is the perturbation of the equilibrium distribution function $F_{0\alpha}$. That is to say, $F_{\alpha}=F_{0\alpha}+f_{\alpha}$.

In this paper, we employ the nonmodal approach to the solution of the Vlasov equation (\ref{5}) with RF fields (\ref{2}) and (\ref{3}), developed in Ref. 8. 
The first step in this approach is the transformation of the velocity $\mathbf{v}$ and the position $\mathbf{r}$ 
coordinates determined in the laboratory frame in Eq. (\ref{5}) to the coordinates $\mathbf{v}_{\alpha}$ and $\mathbf{r}_{\alpha}$, 
\begin{eqnarray}
&\displaystyle 
\mathbf{v}_{\alpha}=\mathbf{v}-\mathbf{V}_{\alpha}\left(\mathbf{r},t \right) , 
\nonumber
\\ 
&\displaystyle
\mathbf{r}_{\alpha}=\mathbf{r}-\mathbf{R}_{\alpha}\left(\mathbf{r},t \right)
= \mathbf{r} -\int\limits^{t} \mathbf{V}_{\alpha}\left(\mathbf{r},t_{1} \right)dt_{1},
\label{7}
\end{eqnarray}
determined in the moving frame of references with a velocity 
$ \mathbf{V}_{\alpha}\left(\mathbf{r},t \right)$. 
With new coordinates, the Vlasov equation for electrons becomes
\begin{eqnarray}
&\displaystyle 
\frac{\partial F_{e}\left(\mathbf{v}_{e},\mathbf{r}_{e},t \right) }{\partial t}
+\mathbf{v}_{e}\frac{\partial F_{e}}{\partial\mathbf{r}_{e}}
-v_{ej}\int\limits^{t}_{t_{0}}\frac{\partial V_{ek}\left( \mathbf{r},t_{1}\right) }{\partial r_{j}}dt_{1}
\frac{\partial F_{e}}{\partial r_{ek}}
\nonumber
\\ 
&\displaystyle
-v_{ej}\frac{\partial V_{ek}\left( \mathbf{r}_{e},t\right)}{\partial r_{ej}}\frac{\partial F_{e}}{\partial v_{ek}}-V_{ej}\left( \mathbf{r}_{e},t\right)
\int\limits^{t}_{t_{0}}\frac{\partial V_{ek}\left( \mathbf{r},t_{1}\right) }{\partial r_{j}}dt_{1}\frac{\partial F_{e}}{\partial r_{ek}}
\nonumber
\\ 
&\displaystyle
+\frac{e}{m_{e}}\left(\nabla \varphi\left( \mathbf{r},t\right)
-\frac{1}{c}\Big[\mathbf{v}_{e}\times\mathbf{B}
_{0}\left(z,t \right) \Big]\right)\cdot\frac{\partial F_{e}}{\partial\mathbf{v}_{e}}
\nonumber
\\ 
&\displaystyle
-\left\{\frac{\partial \mathbf{V}_{e}\left(z, t\right) }{\partial t}+
V_{ez}\left(z, t\right)\frac{\partial \mathbf{V}_{e}\left(z, t\right) }{\partial z}
\right.  
\nonumber
\\ 
&\displaystyle
\left. 
+ \frac{e}{m_{e}}\left(\mathbf{E}_{0y}\left(z,t \right) 
+\frac{1}{c}\Big[\mathbf{V}_{e}\left(z, t\right)\times\mathbf{B}_{0}\left(z, t \right) \Big]\right) 
\right\}
\nonumber
\\ 
&\displaystyle
\cdot\frac{\partial F_{e}\left(\mathbf{v}_{e},\mathbf{r}_{e},t \right)}{\partial \mathbf{v}_{e}}=0.
\label{8}
\end{eqnarray}
Velocity $\mathbf{V}_{e}\left(z, t\right)$ in our approach is determined by the equation vanishing the expression in braces in Eq. (\ref{8}). This 
equation is the Euler equation for the velocity of the ideal fluid in the EM field. 
For the electric field  (\ref{2}) and the magnetic field  (\ref{3}), the equations for $V_{ey}\left(z, t\right)$ and $V_{ez}\left(z, t\right)$ are
\begin{eqnarray}
&\displaystyle 
\frac{\partial V_{ey}\left(z, t\right)}{\partial t}+
V_{ez}\left(z, t\right)\frac{\partial V_{ey}\left(z, t\right)}{\partial z}
\nonumber
\\ 
&\displaystyle
=-\frac{e}{m_{e}}\left(E_{0y}\left(z,t \right) 
+\frac{1}{c}V_{ez}\left(z, t\right)B_{0x}\left(z,t \right)\right),
\label{9}
\\
&\displaystyle 
\frac{\partial V_{ez}\left(z,t\right)}{\partial t}+
V_{ez}\left(z, t\right)\frac{\partial V_{ez}\left(z, t\right)}{\partial z}
\nonumber
\\ 
&\displaystyle
=\frac{e}{m_{e}c}V_{ey}\left(z, t\right)B_{0x}\left(z, t\right).
\label{10}
\end{eqnarray}
In Appendix, we present the detailed solution of these equations for $V_{ey}\left(z,t\right)$ and $V_{ez}\left(z,t\right)$. 

The rest of Eq. (\ref{8}) contains only the spatial derivatives of $\mathbf{V}_{e}\left(z, t\right)$. 
Therefore, Eq. (\ref{8}) in convected variables $\mathbf{v}_{e}$ and $\mathbf{r}_{e}$ has a form as in plasma 
without an EM field for the spatially uniform time-dependent EM field (so-called dipole approximation). 
For the case of the spatially inhomogeneous EM field, Eq. (\ref{8}) becomes very suitable for the investigation of the short scale perturbations with a 
wavelength less then the spatial inhomogeneity of the EM field.
For the considered problem of the skin layer stability, the solution for $F_{e}\left(\mathbf{v}_{e}, \mathbf{r}_{e},t\right)$ may be derived in the form of 
power series in the small parameter $\kappa\delta r_{e}\ll 1$, where $\delta r_{e}$ is the amplitude of the displacement of electron in the RF field.
With velocities $V_{ey}\left(z_{e},t\right)$ and $V_{ez}\left(z_{e},t\right)$, determined by Eqs. (\ref{A14}) and (\ref{A15}), the Vlasov equation 
(\ref{8}) becomes
\begin{eqnarray}
&\displaystyle 
\frac{\partial F_{e}\left(\mathbf{v}_{e},\mathbf{r}_{e}, t\right)}{\partial t}
+\left[v_{ey}+v_{ez}\frac{\omega_{ce}}{\omega_{0}}\sin 
\omega_{0}t e^{-\frac{1}{4}\omega^{2}_{ce}t^{2}}\right]\frac{\partial F_{e}}{\partial y_{e}}
\nonumber
\\ 
&\displaystyle
+v_{ez}\left[1+\left(\frac{1}{2}\omega^{2}_{ce}t^{2}-\frac{1}{4}
\frac{\omega^{2}_{ce}}{\omega_{0}^{2}}\right)e^{-\frac{1}{2}\omega^{2}_{ce}t^{2}}\right]\frac{\partial F_{e}}{\partial z_{e}}
\nonumber
\\ 
&\displaystyle
-\omega_{ce}v_{ez}\left[1-\cos \omega_{0}t e^{-\frac{1}{4}\omega^{2}_{ce}t^{2}}\right]\frac{\partial F_{e}}{\partial v_{ey}}
\nonumber
\\ 
&\displaystyle
+\omega_{ce}\left[v_{ey}+v_{ez}e^{-\frac{1}{2}\omega^{2}_{ce}t^{2}}\left(\omega_{ce}t+\frac{1}{2}\frac{\omega_{ce}}{\omega_{0}}\sin 2\omega_{0}t\right)
\right]\frac{\partial F_{e}}{\partial v_{ey}}
\nonumber
\\ 
&\displaystyle
+\frac{e}{m_{e}}\nabla \varphi\left(\mathbf{r}, t\right)\cdot\frac{\partial F_{e}}{\partial \mathbf{v}_{e}}=0.
\label{11}
\end{eqnarray}

In this equation, $\omega_{ce}$ is the electron cyclotron frequency formed by the magnetic field of the RF wave, which is determined by Eq. (\ref{A4}).
Equation (\ref{11}) is applicable for the treatment of the processes in the skin layer for $\kappa z_{e}\lesssim 1$ as well as those outside of the skin 
layer for 
$\kappa z_{e}> 1$, which occur during a time limited by the interval  $\omega_{0}^{-1}< t < \omega_{ce}^{-1}$.  For the real experimental conditions, this 
time interval is sufficiently wide. For example, for a plasma with electron density $n_{0e}=10^{11}$ cm$^{-3}$ in the electric field  $E_{0y}=1$ V/cm at 
$z_{e}=0$ with a frequency $\omega_{0}=10^{-2}\omega_{pe}= 1.7\times 10^{8}$ s$^{-1}$, and a skin depth $L_{s}=\kappa^{-1}=1.7$ cm, the electron cyclotron 
frequency $\omega_{ce}$ is equal to $6\times 10^{6}\approx 3.5\times 10^{-2}\omega_{0}$. The presence of the small parameter 
$\omega_{ce}/\omega_{0}\ll 1$ in Eq. (\ref{11}) gives a possibility to simplify Eq. (\ref{11}) greatly.

\section{Ion-acoustic instability of the skin layer driven by the accelerated electrons}\label{sec3}

In this section, we present the theory of the ion-acoustic instability in the skin layer, driven by the accelerated electrons.  The growth rate $\gamma$ of 
this 
instability, presented below by Eq. (\ref{20}), is of the order as $\gamma \sim kv_{s}\left(m_{e}/m_{i}\right)^{1/2}$. The IA instability can develop and 
saturate in the skin layer when this growth rate is larger than $\omega_{ci}$. This occurs for the short wavelength perturbations with $k
\rho_{i}> v_{Te}/v_{Ti}$. For the argon plasma, $\left(m_{Ar}/m_{e}\right)^{1/2}=278$, with an 
argon ion temperature  $T_{i}\approx 0.026$ eV and an electron temperature $T_{e}=2$ eV the wave number $k$ of the unstable IA perturbations should be larger than $10^{2}$ cm$^{-1}$.
For the time $t\sim \gamma^{-1}<\omega_{ce}^{-1}$, Eq. (\ref{11}) has a form as in the uniform steady plasma for the electron distribution function $F_{e}$ in 
the electron moving frame and for $F_{i}$ in the ion frame, which in this problem almost coincides with a laboratory frame.  Therefore, we select functions $F_{e0}
\left(\mathbf{v}_{e}\right)$ and $F_{i0}\left(\mathbf{v}_{i}\right)$ as the Maxwellian distributions  in the convective coordinates,
\begin{eqnarray}
&\displaystyle 
F_{\alpha 0}\left(v_{\alpha y},v_{\alpha z}\right) = \frac{n_{0\alpha}}{2\pi v^{2}_{T\alpha} }
\exp\left[-\frac{v^{2}_{\alpha z}+v^{2}_{\alpha y}}{2v^{2}_{Te}}\right]. 
\label{12}
\end{eqnarray}
As displayed in Ref. 8, the single manifestation of the RF wave on the plasma in this case is the accelerated motion of the electrons relative to the practically unmovable ions. The equation for the perturbed electrostatic potential $\varphi\left(\mathbf{k}, t\right)$, Fourier-transformed  in the ion (laboratory) frame, is determined by the Fourier transformation of the Poisson equation in the laboratory frame. This equation for 
the time $t\gg \omega^{-1}_{0}$ has a form

\begin{eqnarray}
&\displaystyle \varphi \left(\mathbf{k}, t\right)+\frac{1}{k^{2}\lambda^{2}_{Di}}\int\limits^{t}_{t_{0}}dt_{1}\varphi \left(\mathbf{k}, t_{1}\right)\frac{d}
{dt_{1}}\left(e^{-\frac{1}{2}k^{2}v^{2}_{Ti}\left(t-t_{1}\right)^{2}}\right)
\nonumber
\\ 
&\displaystyle +\frac{1}{k^{2}\lambda^{2}_{De}}\int\limits^{t}_{t_{0}}dt_{1}\varphi \left(\mathbf{k}, t_{1}\right)e^{-\frac{i}{2}k_{z}a_{ie}
\left(t^{2}-t^{2}_{1}\right)}
\nonumber
\\ 
&\displaystyle
\times\frac{d}{dt_{1}}\left(e^{-\frac{1}{2}k^{2}v^{2}_{Te}\left(t-t_{1}\right)^{2}}\right)=0, 
\label{13}
\end{eqnarray}
where $\lambda_{Di(e)}$ is the ion (electron) Debye length, and $a_{ie}=\omega^{2}_{ce}/2\kappa$. For the adiabatic electrons, the approximation $e^{-\frac{1}
{2}k^{2}v^{2}_{Te}\left(t-t_{1}\right)^{2}}
\approx e^{-k^{2}v^{2}_{Te}t\left(t-t_{1}\right)}$ may be used for  the most fast varying function in the electron term.
For $T_{e}\gg T_{i}$, the electron term which contains the nonmodal time dependence is much less than the ion one. 
Therefore, we are looking for the solution to Eq. (\ref{13}) in the form
\begin{eqnarray}
&\displaystyle \varphi \left(\mathbf{k}, t\right)= \varphi \left(\mathbf{k}\right)e^{-i\omega\left(t\right)t},
\label{14}
\end{eqnarray}
where $\omega\left(\mathbf{k}, t\right)$ slowly changes on the time scale $\sim \omega^{-1}$. 
Then, the partial integration of Eq. (\ref{13}) with $\varphi \left(\mathbf{k}, t\right)$ in a form of Eq. (\ref{14}) gives the following equation 
for $\omega\left(\mathbf{k}, t\right)$,
\begin{eqnarray}
&\displaystyle \varepsilon\left(\mathbf{k}, t\right) \equiv 1+\frac{1}{k^{2}\lambda^{2}_{Di}}\left(1+i\sqrt{\frac{\pi}{2}}z_{i}\left(t\right)
W\left(z_{i}\left(t\right)\right)\right) +
\nonumber
\\ 
&\displaystyle
\frac{1}{k^{2}\lambda^{2}_{De}}\left(1+i\sqrt{\frac{\pi}{2}}z_{e}\left(t\right)
W\left(z_{e}\left(t\right)\right)\right)=Q\left(\mathbf{k}, t, t_{0}\right),
\label{15}
\end{eqnarray}
where $W\left(z\right)=e^{-z^{2}}\left(1 +\left(2i/ \sqrt {\pi } \right)\int\limits_{0}^{z} e^{t^{2}}dt \right)$ is the complex error function, 
$z_{i}\left(t\right)=\omega\left(t\right)/kv_{Ti}$, and $z_{e}\left(t\right)=\left(\omega\left(t\right)-k_{z}a_{ie}t\right)/kv_{Te}$. The function
$Q\left(\mathbf{k}, t, t_{0}\right)$, 
\begin{eqnarray}
&\displaystyle Q\left(\mathbf{k}, t, t_{0}\right)=\frac{1}{k^{2}\lambda^{2}_{Di}}e^{i\omega\left(t\right)\left(t-t_{0}\right)
-\frac{1}{2}k^{2}v^{2}_{Ti}\left(t-t_{0}\right)^{2}}
\nonumber
\\ 
&\displaystyle +\frac{1}{k^{2}\lambda^{2}_{De}}e^{i\left(\omega\left(t\right)-k_{z}a_{ie}t\right)\left(t-t_{0}\right)
-\frac{1}{2}k^{2}v^{2}_{Te}\left(t-t_{0}\right)^{2}},
\label{16}
\end{eqnarray}
determines the input from the $t=t_{0}$ limit of the integration of Eq. (\ref{24}) by parts. For the unstable solutions of Eq. (\ref{15}) for $\omega\left(t
\right)$, function $Q\left(\mathbf{k}, t, t_{0}\right)$ is exponentially small and may be neglected.
The solution of Eq. (\ref{15}) for the IA instability is
\begin{eqnarray}
&\displaystyle \omega\left(t\right)=\omega_{s}\left(k\right)+i\gamma\left(\mathbf{k}, t\right).
\label{17}
\end{eqnarray}
Here, the IA frequency $\omega_{s}\left(k\right)$ is determined as
\begin{eqnarray}
&\displaystyle \omega^{2}_{s}\left(k\right)=\frac{k^{2}v^{2}_{s}}{1+k^{2}\lambda^{2}_{De}},
\label{18}
\end{eqnarray}
where $v_{s}=(T_{e}/m_{i})^{1/2}$ is the ion sound velocity. 
The growth/damping rate of the IA instability is $\gamma\left(\mathbf{k}, t\right)=\gamma_{e}\left(\mathbf{k}, t\right)
+\gamma_{i}\left(\mathbf{k}\right)$, where
\begin{eqnarray}
&\displaystyle \gamma_{i}\left(\mathbf{k}\right)=
-\frac{\omega_{s}\left(k\right)}{\left(1+k^{2}\lambda^{2}_{De}\right)^{3/2}}\sqrt{\frac{\pi}{8}}\left(\frac{T_{e}}{T_{i}}\right)^{3/2}
\nonumber
\\ 
&\displaystyle
\times\exp\left(-\frac{T_{e}}{2T_{i}\left(1+k^{2}\lambda^{2}_{De}\right)}\right)
\label{19}
\end{eqnarray}
stems from the ion Landau damping of the IA waves, and 
\begin{eqnarray}
&\displaystyle \gamma_{e}\left(\mathbf{k}, t\right)
=\sqrt{\frac{\pi m_{e}}{8 m_{i}}}\frac{\omega_{s}\left(k\right)}{\left(1+k^{2}\lambda^{2}_{De}\right)^{3/2}}\left(\frac{k_{z}a_{ie}t}{\omega_{s}
\left(k\right)}-1\right)
\nonumber
\\ 
&\displaystyle
=\gamma_{s}\left(\mathbf{k}, t\right)\left(\frac{k_{z}a_{ie}t}{\omega_{s}
\left(k\right)}-1\right).
\label{20}
\end{eqnarray}

For ICP plasmas with $T_{e} \gg T_{i}$, $\gamma_{i}\left(\mathbf{k}\right)$ is negligible small. The growth rate (\ref{20}) corresponds to the initial linear stage of the IA instability development, at which electrons are uniformly accelerated under the action of the homogeneous electric field. 
In our case, electron acceleration occurs under the action of the effective electric field 
\begin{eqnarray}
&\displaystyle E_{\rm eff}=\frac{1}{2}\frac{e\kappa}{m_{e}}\frac{E^{2}_{0y}}{\omega^{2}_{0}}e^{-2\kappa z_{e}}
= E_{0y}\frac{\omega_{ce}}{2\omega_{0}}e^{-2\kappa z_{e}},
\label{21}
\end{eqnarray}
formed by the ponderomotive force. $E_{\rm eff}=1.75\times 10^{-2}$ V/cm for the numerical sample considered in this paper with $E_{0y}=1$ V/cm at $z_{e}=0$,  $\omega_{0}=10^{-2}\omega_{pe}= 1.7\times 10^{8}$ 
s$^{-1}$, $L_{s}=\kappa^{-1}=1.7$ cm, $\omega_{ce}=6\times 10^{6}$ s$^{-1}$ $\approx 3.5\times 10^{-2}\omega_{0}$.

The instability develops due to the inverse 
electron Landau damping of the IA waves at time $t>t_{th}=\omega_{s}\left(k\right)/
k_{z}a_{ie}$. At that time, the temporal evolution of the IA spectral energy density $W\left(\mathbf{k}, t\right)$,
\begin{eqnarray}
&\displaystyle W\left(\mathbf{k}, t\right)=k^{2} 
\left|\varphi \left(\mathbf{k}\right)\right|^{2}\omega_{s}\left(\mathbf{k}\right)\frac{\partial 
\varepsilon\left(\mathbf{k}, t\right)}{\partial \omega_{s}\left(\mathbf{k}\right)}
\nonumber
\\ 
&\displaystyle
\approx\frac{\omega^{2}_{pi}}{\omega^{2}}k^{2}\frac{1}{4\pi}\left|\varphi \left(\mathbf{k}, t\right)\right|^{2},
\label{22}
\end{eqnarray}
is determined by the equation
\begin{eqnarray}
&\displaystyle \frac{\partial W\left(\mathbf{k}, t\right)}{\partial t}=2\gamma\left(\mathbf{k}, t\right)W\left(\mathbf{k}, t\right)
\label{23}
\end{eqnarray}
and grows with time as $\sim \exp\left(\gamma_{s}k_{z}a_{ei}\left(t-t_{th}\right)^{2}/\omega_{s}\right)$. The  acceleration of the electron current velocity 
and the growth of the IA spectral intensity occur during a limited time until the effect of the ion acoustic turbulence (IAT) on the electron current 
velocity becomes 
negligibly small. At longer times, the temporal evolution of the IAT is determined by the nonlinear 
interaction of the electrons and ions with random electric fields of the IAT\cite{Bychenkov}. 

\section{The nonlinear evolution of the IA instability and the anomalous heating of electrons by the ion-acoustic turbulence}
\label{sec4}
Because the growth rate (\ref{20}) and the growth rate of the conventional IA instability\cite{Akhiezer1} are much less than the IA frequency $\omega_{s}$, 
the nonlinear IAT theory is based on the 
methods of the weak turbulence theory. The conventional IAT theory\cite{Bychenkov} involves the theory of the quasilinear relaxation 
of the electrons on the IA pulsation jointly with the theory of the induced scattering of the IA waves by the ions. 

The quasilinear equation for the ensemble averaged electron distribution 
function $\bar{F}_{e}\left(\mathbf{v}_{e}, \mathbf{r}_{e}, t\right)$ 
in the electron frame
is derived easily from Eq. (\ref{11}) for time $t<\omega_{ce}^{-1}$. It is equal to
\begin{eqnarray}
&\displaystyle \frac{\partial \bar{F}_{e}\left(\mathbf{v}_{e}, \mathbf{r}_{e}, t\right)}{\partial t}=\left\langle \nabla \varphi\left(\mathbf{r},t\right) 
\frac{\partial f_{e}\left(\mathbf{v}_{e}, \mathbf{r}_{e}, t\right)}{\partial\mathbf{v}_{e}}\right\rangle.
\label{24}
\end{eqnarray}
where the angle brackets $\left\langle ...\right\rangle $ indicate the ensemble averaging of the expression in it. Employing the relation
\begin{eqnarray}
&\displaystyle \varphi\left(\mathbf{k}, t\right)= \varphi_{e}\left(\mathbf{k}, t\right)e^{-\frac{i}{2}k_{z}a_{ie}t^{2}}
\label{25}
\end{eqnarray}
between the Fourier transform $\varphi_{e}\left( \mathbf{k}, t\right)$ of the potential $\varphi\left( \mathbf{r}, t\right)$ over 
$\mathbf{r}$  and the Fourier transform $\varphi_{e}\left(\mathbf{k}, t\right)$ of the potential $\varphi_{e}\left( 
\mathbf{r}_{e}, t\right)$ over $\mathbf{r}_{e}$, we derive the quasilinear equation 
\begin{eqnarray}
&\displaystyle 
\frac{\partial \bar{F}_{e}}{\partial t}=\pi\frac{e^{2}}{m_{e}^{2}}\int d\mathbf{k}\mathbf{k}\frac{\partial}{\partial \mathbf{v}_{e}}\left| \varphi
\left(\mathbf{k}\right)\right|^{2}
\nonumber
\\ 
&\displaystyle
\times\delta\left(\omega\left(\mathbf{k}, t\right)-\mathbf{k}\mathbf{v}_{e}-k_{z}a_{ie}t\right)\mathbf{k}\frac{\partial \bar{F}
_{e}}{\partial \mathbf{v}_{e}},
\label{26}
\end{eqnarray}
which determines the temporal evolution of the distribution function $\bar{F}_{e}\left(\mathbf{v}_{e}, \mathbf{r}_{e}, t\right)$ of 
the accelerated electrons under the action of  IAT. By multiplying $\mathbf{v}_{e}$ on Eq. (\ref{26}) and integrating it over $\mathbf{v}_{e}$, 
we derive the equation 
\begin{eqnarray}
&\displaystyle 
\frac{d V_{ez}}{\partial t}=-\nu_{\rm eff}V_{ez}
\nonumber
\\ 
&\displaystyle
=-\frac{1}{n_{0e}m_{e}}\int d\mathbf{k} \frac{k_{z}}{\omega_{s}\left(k\right)}
\gamma_{e}\left(\mathbf{k}, t\right)W\left(\mathbf{k}, t\right),
\label{27}
\end{eqnarray}
which determines slowing down of electrons due to their interactions with the IAT. The temporal evolution of $\nu_{\rm eff}$ depends 
on the temporal evolution of the growth rate $\gamma\left(\mathbf{k}, t\right)$ resulted from the quasilinear distortion of the electron distribution 
function by the IAT and on the temporal evolution of the IAT spectrum $W\left(\mathbf{k}, t\right)$ caused by the induced scattering the of IA waves on ions. 
The theory of the IAT, which simultaneously takes into account both these processes, was developed in Ref.\cite{Bychenkov1}. It was found \cite{Bychenkov} 
that the effects of these processes on the nonlinear evolution of IAT spectrum depends greatly on the value of the applied electric field \cite{Bychenkov}. 
When the electric field $E_{\rm eff}$ is less than $E_{\rm nl}$\cite{Bychenkov,Bychenkov1}, where 
\begin{eqnarray}
&\displaystyle E_{\rm nl}=\frac{m_{e} v_{s}\omega_{pi}}{6\pi |e|}\frac{T_{e}}{T_{i}},
\label{28}
\end{eqnarray}
the quasilinear effects should be accounted for in the balance equation, which includes the growth rate of the IA instability and the 
nonlinear damping rate resulted from 
the induced scattering of IA waves on ions. For the numerical data used in this paper ($T_{e}=2$ eV, $T_{i}=0.026$ eV, $n_{0e}=10^{11}$ cm$^{-3}$), $E_{\rm nl}=0.033$ V/cm which is almost two times larger than $E_{\rm eff}=1.75\times 10^{-2}$\,V/cm of our sample. 
In this case, the level $W=\int W\left(\mathbf{k}\right)d\mathbf{k}$ of the total energy density of the IAT determined by this process does not depend on the 
magnitude of the applied electric field $E_{0}$ and was estimated\cite{Bychenkov} by the expression
\begin{eqnarray}
&\displaystyle 
\frac{W}{n_{0e}T_{e}} \sim 0.1\frac{\omega_{pi}}{\omega_{pe}}\frac{\lambda^{2}_{De}}{\lambda^{2}_{Di}},
\label{29}
\end{eqnarray}
where $\omega_{pi}$ is the ion plasma frequency. In the approach employed in our paper, the ion dynamics in the skin layer is not affected by 
the EM field. Therefore, the theory of the induced scattering of the IA waves on ions developed in Ref. 14 
is completely applicable to the IA instability driven by the ponderomotive current for the time $t< \omega_{ce}^{-1}$, considered here. For the numerical parameters presented above the level (\ref{29}) is estimated as
\begin{eqnarray}
&\displaystyle 
\frac{W}{n_{0e}T_{e}} \sim 2.8\times 10^{-2}.
\label{30}
\end{eqnarray}
In estimate (\ref{30}), it should be accounted for that the directed accelerated velocity $V_{ez}\left(t
\right)$ is slowed down to the almost threshold velocity of the order of IA velocity\cite{Bychenkov} at the nonlinearly established steady state of the IAT. 
On this level, the effective electron collision frequency with the IAT, $\nu_{\rm eff}$, determined by Eq. (\ref{27}), is estimated as 
\begin{eqnarray}
&\displaystyle \nu_{\rm eff}\sim\frac{W}{n_{0e}m_{e}V_{ez}}\frac{\gamma_{s}}{v_{s}}\sim 2.8\times 10^{-2}\omega_{s}\left(\frac{m_{i}}{m_{e}}\right)^{1/2}.
\label{31}
\end{eqnarray}
Because the induced scattering of ions redistributes the spectral maximum of the IAT to the longer IA waves, we use a value $k=10^{2}$ cm$^{-1}$ discussed above for the IA waves in the skin layer for the estimation of $\nu_{\rm eff}$. 
We found from Eq. (\ref{30}) that the magnitude of $\nu_{\rm eff}\sim 1.7\times 
10^{8}$ s$^{-1}$ for this case is of the order of the frequency $\omega_{0}=1.8\times 10^{8}$ s$^{-1}$ used in our estimates.
At this case of the weak electric field, $E_{\rm eff} < E_{\rm nl}$, a quasi-stationary state for the IAT is established  mainly  due to the quasilinear 
relaxation of the electron distribution function. Ohm's law for the electron current density $j$, derived for this case\cite{Bychenkov, Galeev}, 
\begin{eqnarray}
&\displaystyle j\simeq 2.14|e|n_{0e}v_{s}=\sigma_{A} E_{\rm eff},
\label{32}
\end{eqnarray}
predicts the dependence of $\sigma_{A}\sim E^{-1}_{\rm eff}$ for the anomalous conductivity $\sigma_{A}$.
By multiplying $m_{e}\mathbf{v}^{2}_{e}/2$ on Eq. (\ref{26}) and integrating it over $\mathbf{v}_{e}$, we derive the equation 
\begin{eqnarray}
&\displaystyle 
n_{e0}\frac{d T_{e}}{\partial t}=\int d\mathbf{k} \frac{\left(k_{z}V_{ez}\left(t\right)-\omega_{s}\left(k\right)\right)}{\omega_{s}\left(k\right)}
\gamma_{e}\left(\mathbf{k}, t\right)W\left(\mathbf{k}\right)
\nonumber
\\ 
&\displaystyle
\sim \gamma_{s}\left(\mathbf{k}_{0}\right)\frac{W}{n_{0e}T_{e}}T_{e},
\label{33}
\end{eqnarray}
which determines the turbulent heating rate $\nu_{Te}\sim \gamma W/n_{e0}T_{e}$ of the electrons due to their interaction with the IAT. 
For the numerical data, used above, $\nu_{Te}\sim 2.8 \times 10^{-2}\gamma_{s}$.

When the applied electric field $E_{\rm eff}$ is above $E_{\rm nl}$, quasilinear effects are weak and the dominant nonlinear process is the induced 
scattering of IA waves on ions. In this case, the steady state level of the IAT\cite{Bychenkov},
\begin{eqnarray}
&\displaystyle 
\frac{W}{n_{0e}T_{e}} \sim 0.1\frac{\omega_{pi}}{\omega_{pe}}\frac{T_{e}}{T_{i}}\sqrt{\frac{E_{\rm eff}}{E_{\rm nl}}},
\label{34}
\end{eqnarray}
grows with $E_{\rm eff}$ growth. The Ohm's law for electron current density $j$ for this case, 
\begin{eqnarray}
&\displaystyle 
j=\frac{4.48}{\pi}|e|n_{0e}v_{s}\sqrt{\frac{E_{\rm eff}}{E_{\rm nl}}}=|e|U_{0}n_{0e},
\label{35}
\end{eqnarray}
and the anomalous conductivity $\sigma_{A}$, 
\cite{Bychenkov, Bychenkov1}
\begin{eqnarray}
&\displaystyle 
\sigma_{A}\simeq 0.4\omega_{pe}\frac{\lambda_{Di}}{\lambda_{De}}\left(\frac{8\pi n_{e0}T_{e}}{E_{\rm eff}^{2}}\right)^{1/4} \sim E_{\rm eff}^{-1/2}.
\label{36}
\end{eqnarray}
were derived in Ref. 14. 
The effective collision frequency $\nu_{\rm eff}$  corresponding to this case is given by the Sagdeev equation
\cite{Sagdeev},
\begin{eqnarray}
&\displaystyle 
\nu_{\rm eff}=2.5\times 10^{-2}\omega_{pi}\frac{U_{0}}{v_{s}}\frac{T_{e}}{T_{i}},
\label{37}
\end{eqnarray}
where $U_{0}$ is determined from Eq. (\ref{36}).  The regime with $E_{\rm eff}>E_{\rm nl}$ occurs for $E_{0y}=1.3$ V/cm with the same other parameters considered 
above. For this electric field  $\omega_{ce}/\omega_{0} =4.6\times 10^{-2}$ and $E_{\rm eff}\approx 5.89\times 10^{-2}$ V/cm $> E_{\rm nl}=3.3\times 10^{-2}$ V/cm, 
that gives
\begin{eqnarray}
&\displaystyle 
\frac{W}{n_{0e}T_{e}} \sim 0.05,
\label{38}
\end{eqnarray}
$U_{0}=2.57 v_{s}$ and $\nu_{\rm eff}=8.35\times 10^{8}\,{\text s}^{-1}>\omega_{0}=1.8\times 10^{8}\,{\text s}^{-1}$.

The derived estimates for $\nu_{\rm eff}$ reveal that 
the effective electron collision frequency with the IAT pulsations is of the order of the RF driving frequency and is much larger than the electron-ion and 
electron-neutral collision frequencies in the mTorr range of gas pressure for the considered numerical parameters corresponding to the experimental 
conditions of ICP sources.

\section{Conclusions}\label{sec5}
In this paper, we present the theory of IA instability of the skin layer of ICP sources driven by the accelerated electrons, which move relative to ions 
under the ponderomotive force. This theory reveals that on the linear stage of the IA instability, driven by the steady electric field,
always develops as the nonmodal instability with the growth rate growing with time. 

At the finite time interval $\delta t< \omega_{ce}^{-1}$, the analysis of the nonlinear stage of the IA instability driven by the ponderomotive force is 
similar to the analysis of the IA instability and the IAT driven by the steady electric field. The accelerated electron velocity in the steady electric field 
decelerates due to the scattering of electrons by the IAT. This velocity approaches a particular steady value or continues to be accelerating\cite{Sagdeev} 
depending on the balance of the nonmodal growth with nonlinear processes: the quasilinear relaxation of the electron distribution function and the induced 
scattering of ions. 
The effect of these processes on the nonlinear evolution of the IAT depends on the relative value of the driving electric field with respect to $E_{\rm nl}$. 
The acting electric field in the skin layer application of the IAT theory is the effective field $E_{\rm eff}$ determined by Eq. (\ref{21}), which is 
strongly inhomogeneous across the skin layer in ICPs. It is found that the effective electron collision frequency $\nu_{\rm eff}$ with the IAT is of the 
order of or is larger than $\omega_{0}$ at all considered regimes of the IAT evolution. The derived results prove that the development of the IAT in the skin 
layer and scattering of electrons by IAT are the primary channels of the nonlinear absorption of the RF wave energy in the skin layer. 
This result is also valid for the case of the strong RF field considered in Ref. 8, 
for which the oscillatory velocity in the skin layer is larger than the electron thermal velocity, and the IA instability develops in the bulk of plasma past 
the skin layer.

\begin{acknowledgments}
This work was supported by National R\&D Program through the National Research Foundation of Korea (NRF) funded by the Ministry of Education, Science and 
Technology (Grants Nos. NRF--2018R1D1A1B07050372 and NRF--2019R1A2C1088518) and BK21 PLUS, the Creative Human Resource Education and Research Programs for 
ICT Convergence in the 4th Industrial Revolution.
\end{acknowledgments}

\bigskip
{\bf DATA AVAILABILITY}

\bigskip
The data that support the findings of this study are available from the corresponding author upon reasonable request.

\appendix
\section{{Solutions to Eqs. (\ref{9}) and (\ref{10}) for $V_{ey}$ and $V_{ez}$}}
Here, we present the solutions to Eqs. (\ref{9}) and (\ref{10}) for $V_{ey}$ and $V_{ez}$, alternative to those presented in Ref. 8. These 
solutions do not require the usually employed assumption\cite{Schmidt} in the calculation of the electron velocity in EM wave that the 
force by the RF electric field acting on electrons in the EM wave prevails over the Lorentz force by the RF magnetic field. 

With new variables $z_{e}$ and $t'$ determined by the relations\cite{Davidson, Mikhailenko3}
\begin{eqnarray}
&\displaystyle 
z= z_{e}+\int\limits^{t'}_{0}V_{ez}\left(z_{e},t'_{1} \right) dt'_{1}, \quad t=t'.
\label{A1}
\end{eqnarray}
Eqs. (\ref{9}) and (\ref{10}) becomes
\begin{eqnarray}
&\displaystyle 
\frac{\partial V_{ey}\left( z_{e},t'\right)}{\partial t'} =
-\frac{\omega_{0}\omega_{ce}}{\kappa}e^{-\kappa\int\limits^{t'}_{0}V_{ez}\left(z_{e},t'_{1} \right)dt'_{1}}\sin \omega_{0}t'
\nonumber
\\ 
&\displaystyle
-\omega_{ce}\cos\omega_{0}t'e^{-\kappa\int\limits^{t'}_{0}V_{ez}\left(z_{e},t'_{1} \right)dt'_{1}}V_{ez}\left(z_{e},t' \right), 
\label{A2}
\end{eqnarray}
\begin{eqnarray}
&\displaystyle 
\frac{\partial V_{ez}\left( z_{e},t'\right) }{\partial t'} 
=\omega_{ce}e^{-\kappa \int\limits^{t'}_{0}V_{ez}\left(z_{e},t'_{1} \right) dt'_{1}}
\nonumber
\\ 
&\displaystyle
\times V_{ey}\left(z_{e},t' \right)\cos\omega_{0}t',
\label{A3}
\end{eqnarray}
where 
\begin{eqnarray}
&\displaystyle \omega_{ce}=\frac{e\kappa E_{0y} e^{-\kappa z_{e}}}{m_{e}\omega_{0}}
\label{A4}
\end{eqnarray}
is the electron cyclotron frequency formed by the RF magnetic field  (\ref{3}) at $z_{e}$. Now we derive the approximate solution to Eqs. (\ref{A2}) and (\ref{A3}) for the finite time interval, at which 
\begin{eqnarray}
&\displaystyle \left|\kappa\int\limits^{t'}_{0}V_{ez}\left(z_{e},t'_{1} \right)dt'_{1}\right|\ll 1.
\label{A5}
\end{eqnarray}
In the zero approximation, Eqs. (\ref{A2}) and (\ref{A3}) are reduced to the ordinary differential equation
\begin{eqnarray}
&\displaystyle \frac{d U_{e}}{d t'}-i\omega_{ce}\cos \omega_{0}t \,U_{e}=-\frac{\omega_{0}\omega_{ce}}{\kappa}\sin \omega_{0}t,
\label{A6}
\end{eqnarray}
in which variable $z_{e}$ becomes a parameter, and
\begin{eqnarray}
&\displaystyle U_{e}=U_{e}\left(z_{e},t\right)= V_{ey}\left(z_{e},t\right)+iV_{ez}\left(z_{e},t\right).
\label{A7}
\end{eqnarray}
The solution to Eq. (\ref{A6}) for the initial value $U_{e}\left(z_{e},t=0\right)=0$,
\begin{eqnarray}
&\displaystyle U_{e}\left(z_{e},t\right)= -\frac{\omega_{0}\omega_{ce}}{\kappa}
\exp\left( i\frac{\omega_{ce}}{\omega_{0}}\sin \omega_{0}t\right) 
\nonumber
\\ 
&\displaystyle
\times
\int\limits^{t}_{0}dt_{1}\sin \omega_{0}t_{1}
\exp\left( -i\frac{\omega_{ce}}{\omega_{0}}\sin \omega_{0}t_{1}\right) ,
\label{A8}
\end{eqnarray}
is simply presented in the explicit, however cumbersome, form for any values of the $\omega_{ci}/\omega_{0}$ ratio. 
The focus of our paper is on the weak high-frequency RF field for which $\omega_{ce}/\omega_{0}\ll 1$. By using the approximation 
for the exponent in Eq. (\ref{A8}), 
\begin{eqnarray}
&\displaystyle
\exp\left(\pm i
\frac{\omega_{ce}}{\omega_{0}}\sin \omega_{0}t\right)\approx 1\pm i\frac{\omega_{ce}}{\omega_{0}}\sin \omega_{0}t
\label{A9}
\end{eqnarray} 
which is valid for any values of $\omega_{0}t$, we derive the approximate solution for $U_{e}\left(z_{e},t\right)$  from Eq. (\ref{A8}) in the form
\begin{eqnarray}
&\displaystyle U_{e}\left(z_{e},t\right)= \frac{\omega_{ce}}{\kappa}\cos \omega_{0}t+i\frac{\omega^{2}_{ce}t}{2\kappa}
+i\frac{\omega^{2}_{ce}}{4\omega_{0}\kappa}\sin 2\omega_{0}t.
\label{A10}
\end{eqnarray}
Equation (\ref{A10}) gives
\begin{eqnarray}
&\displaystyle V_{ey}\left(z_{e},t\right)= \mathrm{Re}\,U_{e}\left(z_{e},t\right)= \frac{\omega_{ce}}{\kappa}\cos\omega_{0}t
\label{A11}
\end{eqnarray}
and
\begin{eqnarray}
&\displaystyle 
V_{ez}\left(z_{e},t\right)=\mathrm{Im}\,U_{e}\left(z_{e},t\right)
\nonumber
\\ 
&\displaystyle
=\frac{\omega^{2}_{ce}t}{2\kappa}+\frac{\omega^{2}_{ce}}{4\omega_{0}\kappa}\sin 2\omega_{0}t.
\label{A12}
\end{eqnarray}
Solutions (\ref{A11}) and (\ref{A12}) are identical to solutions for $V_{ey}$ and $V_{ez}$, derived in Ref. 8, where other procedure 
was developed by the iterative solution of Eqs. (\ref{9}) and (\ref{10}). It was based on the calculation of the ponderomotive motion
of an electron in the spatially inhomogeneous electromagnetic field, assuming that the force by the RF electric field acting on electrons in the skin layer 
prevails over the Lorentz force by the RF magnetic field. The procedure developed in Ref. 8 used small parameter $\kappa\xi_{e}\ll 1$ 
where $\xi_{e}=eE_{0y}\left(z_{e}\right)/m_{e}\omega^{2}_{0}$ is the amplitude of the displacement of an electron along the coordinate $y$.  This parameter is  identically equal  to ${\omega_{ce}}/{\omega_{0}}$. For the time at which $\omega_{0}t\gg 1$,  
\begin{eqnarray}
&\displaystyle V_{ez}\left(z_{e},t\right)\approx \frac{\omega_{ce}^{2}}{2\kappa}t
\label{A13}
\end{eqnarray}
and 
\begin{eqnarray}
&\displaystyle -\kappa\int\limits^{t'}_{0}V_{ez}\left(z_{e},t'_{1} \right)dt_{1}'=-\frac{1}{4}\omega_{ce}^{2}t^{2}.
\label{A14}
\end{eqnarray}
It follows from Eq. (\ref{A5}) that solutions (\ref{A11}) and (\ref{A12}) are valid for the time $t<\omega_{ce}^{-1}$. 

By employing the method of successive approximations  to the solution of the nonlinear Eqs. (\ref{A2}) and (\ref{A3}) with $V_{ez}\left(z_{e},t \right)$
determined by Eq. (\ref{A12}) as the initial approximation, we obtain  the following solutions to these equations for $V_{ey}\left(z_{e},t \right)$,
\begin{eqnarray}
&\displaystyle V_{ey}\left(z_{e},t\right)= \frac{\omega_{ce}}{\kappa}e^{-\frac{1}{4}\omega^{2}_{ce}t^{2}}\cos \omega_{0}t,
\label{A15}
\end{eqnarray}
and for $V_{ez}\left(z_{e},t\right)$,
\begin{eqnarray}
&\displaystyle V_{ez}\left(z_{e},t\right)=\frac{\omega^{2}_{ce}t}{2\kappa}e^{-\frac{1}{2}\omega^{2}_{ce}t^{2}}
\nonumber
\\ 
&\displaystyle
+\frac{\omega^{2}_{ce}}{4\omega_{0}\kappa}e^{-\frac{1}{2}\omega^{2}_{ce}t^{2}}\sin 2\omega_{0}t +O\left(\frac{\omega^{2}_{ce}}{\omega^{2}_{0}}\right),
\label{A16}
\end{eqnarray}
which are valid for the time $t\gg \omega^{-1}_{0}$. It follows from Eq. (\ref{A16}) that the maximum of the accelerating velocity 
$V_{ez}\left(z_{e},t\right)$ attains for $t_{\ast}=\sqrt{2}/\omega_{ce}$ at which $\omega^{2}_{ce}t^{2}=2$. At the time $t>2t_{\ast}$, an electron which was 
in $z_{e}=0$ at the time $t=0$ will cover the distance of the order of the skin depth. Therefore, the time admissible for the development and 
saturation of any instability in the skin layer driven by the accelerated electron current in the case of the high operation frequency for which $\omega_{0}
\gg \omega_{ce}$ is limited by time $t\lesssim t_{\ast}\sim \omega_{ce}^{-1}$. 

In the case of the low frequency $\omega_{0}$, large amplitude RF wave, the electron cyclotron frequency $\omega_{ce}$ may be larger than $\omega_{0}$
\cite{Cohen1, Cohen2}. The approximate solution to Eq. (\ref{A8}) can be derived in this case for the limited time interval $t\ll \omega_{0}^{-1}$, 
by employing the simplest approximation $\sin\omega_{0}t \approx \omega_{0}t \ll 1$. In this time interval, the approximate solution to Eq. (\ref{A8}) is
\begin{eqnarray}
&\displaystyle U_{e}\left(z_{e},t\right)= -i\frac{\omega^{2}_{0}t}{\kappa}-\frac{\omega^{2}_{0}}{\kappa \omega_{ce}\left(z_{e}\right)}
\nonumber
\\ 
&\displaystyle
+\frac{\omega^{2}_{0}}{\kappa\omega_{ce}\left(z_{e}\right)}e^{i\omega_{ce}\left(z_{e}\right)t}.
\label{A17}
\end{eqnarray}
It follows from Eq. (\ref{A17}) that 
\begin{eqnarray}
&\displaystyle V_{ey}\left(z_{e},t\right)= \frac{\omega^{2}_{0}}{\kappa \omega_{ce}\left(z_{e}\right)}\left(1-\cos\omega_{ce}\left(z_{e}\right)t\right) 
\label{A18}
\end{eqnarray}
and 
\begin{eqnarray}
&\displaystyle V_{ez}\left(z_{e},t\right)= -\frac{\omega^{2}_{0}t}{\kappa}+\frac{\omega^{2}_{0}}{\kappa \omega_{ce}\left(z_{e}\right)}
\sin \omega_{ce}\left(z_{e}\right)t 
\nonumber
\\ 
&\displaystyle
\approx -\frac{\omega^{2}_{0}t}{\kappa}.
\label{A19}
\end{eqnarray}
Note, that condition (\ref{A5}) with velocity $V_{ez}\left(z_{e},t\right)$ determined by Eq. (\ref{A19}) is valid at time $t\ll \omega_{0}^{-1}$.
The derived solution (\ref{A17}) can be easily improved by using the expansion $\sin \omega_{0}t\approx \omega_{0}t 
-\frac{1}{6}\left(\omega_{0}t\right)^{3}$. In this case, the solution for $V_{ez}\left(z_{e},t\right)$ becomes equal to
\begin{eqnarray}
&\displaystyle V_{ez}\left(z_{e},t\right)\approx -\frac{\omega_{0}}{\kappa}\left(1-3\frac{\omega^{2}_{0}}{\omega^{2}_{ce}}\right)\omega_{0}t.
\label{A20}
\end{eqnarray}


\begin{thebibliography}{}

\bibitem{Lieberman}M.~A.~Lieberman and A.~I.~Lichtenberg, \textit{Principles of Plasma Discharges
and Materials Processing}. 2nd ed. Wiley, New York, 2005.

\bibitem{Weibel} E.~S.~Weibel, Phys. Fluids {\bf 10}, 741 (1967).

\bibitem{Kolobov}V.~I.~Kolobov, D.~J.~Economou, Plasma Sources Sci. Technol. {\bf 6}, R1 (1997).

\bibitem{Alexandrov} A.~F.~Alexandrov, L.~S.~Bogdankevich, A.~A.~Rukhadze, \textit {Principles of Plasma Electrodynamics}.~Springer-Verlag, Berlin, 1984. 

\bibitem{Silin}V.~P.~Silin, Zh. Eksp. Teor. Fiz. {\bf 48}, 1679 (1965); Sov. Phys. JETP {\bf 21}, 1127 (1965).

\bibitem{Porkolab} M.~Porkolab, Nuclear Fusion 1978 {\bf 18},367 (1978).

\bibitem{Akhiezer} A.~I.~Akhiezer, V.~S.~Mikhailenko, K.~N.~Stepanov, Physics Letters A {\bf 245}, 117 (1998).

\bibitem{Mikhailenko3}V.~V.~Mikhailenko, V.~S.~Mikhailenko, H.~J.~Lee, Phys. Plasmas {\bf 27}, 072102 (2020).

\bibitem{Akhiezer1} A.~I. Akhiezer, I.~A.~Akhiezer, R.~V.~Polovin, A.~G.~Sitenko, and K.~N.~Stepanov, 
\textit {Plasma Electrodynamics}.~Pergamon, New York, 1975.

\bibitem{Mikhailenko2} V.~V.~Mikhailenko, V.~S.~Mikhailenko, H.~J.~Lee, Physics of Plasmas {\bf 25}, 012902 (2018).

\bibitem{Davidson} R.~C.~Davidson, \textit{Methods in Nonlinear Plasma Theory}.~ Academic, New York, 1972.

\bibitem{Schmidt} G.~Schmidt, \textit{Physics of High Temperature Plasmas}.  ~Academic, New York, 1979.

\bibitem{Cohen1} R.~H.~Cohen, T.~D.~Rognlien, Plasma Sources Sci. Technol. {\bf 5}, 442 (1996).

\bibitem{Cohen2} R.~H.~Cohen, T.~D.~Rognlien, Phys. Plasmas. {\bf 3}, 1839 (1996).

\bibitem{Galeev} A.~A.~Galeev, R.~Z.~Sagdeev. Review of Plasma Physics. vol. 7. Springer Science+Business Media, New York, 1979. 

\bibitem{Bychenkov} V.~Yu.~Bychenkov, V.~P.~Silin, and S.~A.~Uryupin, Phys. Reports {\bf 164}, 119 (1988).

\bibitem{Bychenkov1} V.~Yu.~Bychenkov, V.~P.~Silin, Sov Phys. JETP {\bf 55}, 1086 (1982).

\bibitem{Sagdeev} A.~A.~Galeev, R.~Z.~Sagdeev, "Current instabilities and anomalous resistivity of plasma," in Basic Plasma Physics: Selected Chapters, Handbook of Plasma Physics, edited by A. A. Galeev and R. N. Sudan (North-Holland Publishing Company, Amsterdam, 1984), Vol. II, pp. 271–303.

\end{thebibliography}
\end{document}